\documentclass[%
 aip,
 rsi,%
 amsmath,amssymb,
reprint,%
]{revtex4-1}

\usepackage{graphicx}% Include figure files
\usepackage{dcolumn}% Align table columns on decimal point
\usepackage{bm}% bold math
\begin{document}

\preprint{AIP/123-QED}

\title[Sample title]{Precise Measurement of a Magnetic Field Generated by the Electromagnetic Flux Compression Technique}% Force line breaks with \\

\author{D. Nakamura}
 \email{dnakamura@issp.u-tokyo.ac.jp}

\author{H. Sawabe}%

\author{Y. H. Matsuda}%
  
\author{S. Takeyama}
\affiliation{ 
Institute for Solid State Physics, University of Tokyo, 5-1-5, Kashiwanoha, Kashiwa, Chiba 277-8581,
Japan%\\This line break forced with \textbackslash\textbackslash
}%

\date{\today}% It is always \today, today,
             %  but any date may be explicitly specified

\begin{abstract}
The precision of the values of a magnetic field generated by electromagnetic flux compression was investigated in ultra-high magnetic fields of up to 700 T.
In an attempt to calibrate the magnetic field measured by pickup coils, precise Faraday rotation (FR) measurements were conducted on optical (quartz and crown) glasses.
A discernible "turn-around" phenomenon was observed in the FR signal as well as the pickup coils before the end of a liner implosion.
We found that the magnetic field measured by pickup coils should be corrected by taking into account the high-frequency response of the signal transmission line.
Near the peak magnetic field, however, the pickup coils failed to provide reliable values, leaving the FR measurement as the only method to precisely measure an extremely high magnetic fields.
\end{abstract}

\pacs{41.20.Jb,07.55.Db,78.20.Ls}
%41.20.Jb Electromagnetic wave propagation; radiowave propagation
%07.55.Db Generation of magnetic fields; magnets
%78.20.Ls	Magneto-optical effects
% PACS, the Physics and Astronomy
                             % Classification Scheme.
\keywords{electromagnetic flux compression, ultra-high magnetic field, metrology, Faraday rotation}%Use showkeys class option if keyword
                              %display desired
\maketitle

\section{\label{sec:level1}Introduction}
The application of high magnetic fields is widespread in fields such as physics, chemistry, biology, and medicine.
In particular, in the region of ultra-high magnetic fields, the Zeeman splitting energy and the cyclotron resonance energy (of the quantized orbital motion of free electrons) can exceed competing energy scales such as thermal fluctuations, and we can access the quantum limit of materials, even at a room temperature.
The demand for high magnetic fields in solid state physics applications, is as a result, rapidly growing.
An ultra-high magnetic field (above 100 T) can only be achieved by a destructive type of magnet due to Maxwell stresses exceeding 40,000 kg/cm$^2$.\cite{MiuraHerlach1985,Herlach1999RPP,MiuraHerlach2003}
%destructive magnets
Pulsed magnetic fields are generated by destruction of magnets over a timescale of microseconds.
The flux compression and single-turn coil techniques are currently the only available methods for the generation of ultra-high magnetic fields.
In the single-turn coil technique,\cite{Miura2001PhysB} a magnetic field of up to 300 T is generated by a mega-ampere current injected into a single-folded coil.
In the flux compression technique, a seed field is injected into the main coil and compressed by the implosion of a metallic cylinder called a "liner."
%The flux compression method utilizing the dynamics of imploding liner.
The explosive-driven flux compression technique \cite{Fowler1960,Bykov2001PB} uses chemical explosives for accelerating the implosive liner, and can access magnetic fields of around 1000 T.
However, a single experiment requires substantial preparation, and the destructive nature of this technique causes difficulties with respect to reproducibility and controllability.
As a result, the application of this technique has lagged behind the single-turn coil technique with regards to experimental solid state physics.

The electromagnetic flux compression (EMFC) method, on the other hand, first developed by Cnare \cite{Cnare1966} in the 1960's, is more suitable for application to experimental solid state physics.
Figure 1 shows a schematic of a coil used for the EMFC method.
In the EMFC method, the liner implosion is caused by an electromagnetic (repulsive) force from the single-turn (primary) coil at the moment when a huge current is injected into this coil.
Regardless of the destructive nature of this method, it has the advantage of precise controllability, and is more suitable for indoor experiments than any chemical explosive methods.
Recently, we reported on the generation of magnetic fields of up to 730 T: the world's highest ever magnetic field generated in an indoor laboratory.
This was achieved by a newly designed copper-lined primary coil (shown in Fig. 1),\cite{Takeyama2010,Takeyama2011} and has already been applied to several solid state measurements.
For example, interesting results have been obtained for the high $T_c$ cuprate superconductors,\cite{Sekitani2004} carbon nanotubes,\cite{Takeyama2011CNT} and frustrated magnets.\cite{Miyata2011}

\begin{figure}[t]
\includegraphics[width=7cm]{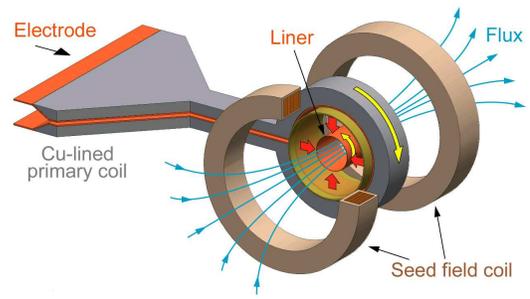}
\caption{\label{fig:1} (color online) Schematic of the EMFC magnet, composed of a copper-lined primary coil with a copper liner inside, and a pair of seed field coils.}
\end{figure}

%metrology in HMF
Regardless of the progress of the EMFC technique that has been made over the past 40 years, less attention has been paid so far to the precision of the measured magnetic field.
As the maximum magnetic field and rate of change of the magnetic flux have increased, the accuracy of the measurement of the magnetic field in these experiments has become more important.
In a pulsed operation the magnetic field is determined from the voltage induced in a pickup coil inserted into the center of the magnet coil, whereas the pickup coil is set at the center of imploding liner in EMFC experiments.
When the EMFC technique is used for the purpose of solid state physics measurements, the pickup coil is the only method available as a magnetic field probe due to the limited space (the diameter of the final space inside the imploding liner is less than 6 mm).\cite{Takeyama2011}
However, a pickup coil is not necessarily the ideal probe for an accurate determination of ultra-high magnetic fields.
The difficulties involved with the evaluation of a precise magnetic field using pickup coils are the following:
\begin{itemize}
 \item The precision of the pickup coil dimension degrades with reducing size. 
 \item The electrical insulation breaks down easily for induced voltages as high as a few kV. 
 \item The wire used in the pickup coils heats up as a result of the current induced from the huge $d\phi/dt$ ($\phi$: magnetic flux). 
 \item The high frequency characteristics of the pickup signal are transferred to the measurement instruments.
\end{itemize}
In addition, a fictitious signal is often observed when the pickup coil is damaged or destroyed during the increase of magnetic field, which can lead to misinterpretation of the experimental results.
Accordingly, it is important to pursue a precise and reliable calibration of the pickup coil for measurement of the magnetic field, by some other method.

%magneto-optical method
To deal with this issue, we have attempted to employ a magneto-optical method as a probe for the magnetic field, which has some advantages compared with the pickup coil.
%merit of the optical measurement
For example, optical measurements do not require a metal lead wire around the sample, and are thus less influenced by the effect of electromagnetic noise as well as also escaping the breakdown of electrical insulation.
Furthermore, complex electric components such as an integrator circuit are avoided, which simplifies the calibration of the magnetic field, and reduces possible sources of error.
Therefore, the optical measurement is expected to be quite useful for the precise evaluation of ultra-high magnetic fields.

We adapted the Faraday rotation (FR) method for optical measurement of the magnetic field.
Optically transparent materials generally exhibit a linear response of the FR angle ($\theta_F$), with respect to an external magnetic field, as described by: 
\begin{equation}
 \theta_F = v LB, 
\end{equation}
where $v$ is the Verdet constant of the material, $L$ is the sample length, and $B$ is the external magnetic field.
By measuring the $\theta_F$ of a material whose $v$ is known, if $L$ is accurately measured, the magnetic field can be precisely evaluated.
In addition, the pickup coil can be calibrated precisely by means of a simultaneous measurement.

Another candidate for the calibration of ultra-high magnetic fields is a resonant type of measurement.\cite{Garn1966,Kido1982APL}
For example, Kido {\it et al.} used the ESR (Electron Spin Resonance) signal of Ruby, whose ESR peak appears at 91.0 T for a laser wavelength of 118.65 $\mu$m.\cite{Kido1982APL}
However, despite its high precision, this method is only applicable to a discrete point. 

\begin{figure*}[thbp]
\includegraphics[width=16cm]{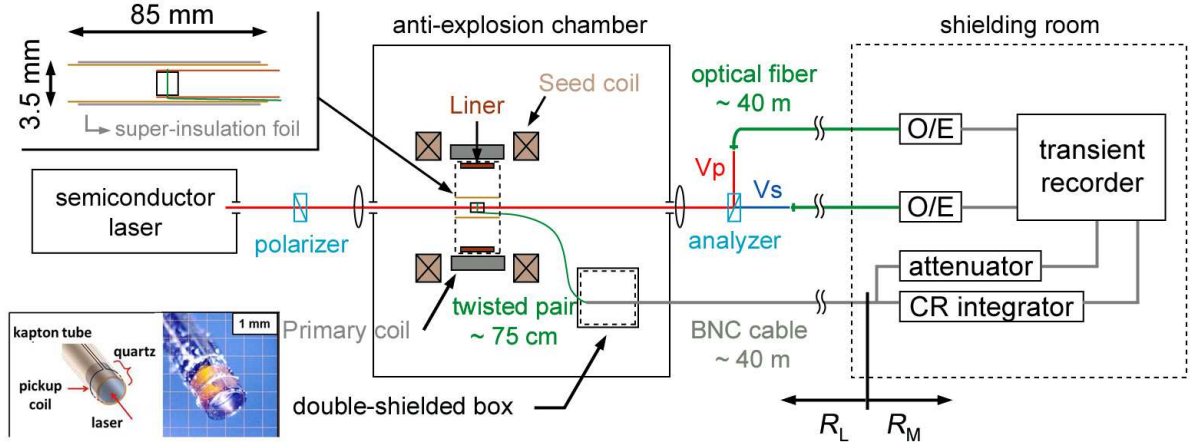}
\caption{\label{fig:2} (color online) Optics setup for the FR measurements with the EMFC coil and the sample holder set inside an anti-explosion chamber. Inset (bottom left) shows the sample holder, a 2 mm-thick quartz rod with a pickup coil wound around the rod.}
\end{figure*}

The measurement of magnetic field by the FR method has been previously applied to the flux compression technique,\cite{George1965,Alikhanov1968,Garn1968,Kido1976,Wessel1986} however, to the best of our knowledge, there has been no detailed study of the precise evaluation of the magnetic field above 300 T.
In this article, we analyze the results of FR measurements in ultra-high magnetic fields (up to $\sim$ 700 T) for two types of optical glass, and compare this to measurements of the magnetic field from pickup coils.
At the final stage of the flux compression, a peak structure due to leakage of the magnetic flux from the imploding liner is observed, which is known as the "turn-around" phenomenon.\cite{Takeyama2011}
The observation of this phenomenon is regarded as an important indication of measuring up to the maximum magnetic field.
An optical FR transmission signal was captured up to this final stage of the flux compression, as indicated by the observation of this turn-around structure.

\section{\label{sec:level2}Experimental Method}

Details of the experimental setup for EMFC are described in reference \cite{Takeyama2011}; the coil was set in an anti-explosion chamber for explosive experiments. The overall experimental setup is illustrated in Fig. 2.
The FR of an optical glass rod was measured simultaneously to the signal of a pickup coil in the manner shown in the inset of Fig. 2.
The sample probe was inserted into the center position of the pickup coil.
For FR measurements we used fused quartz and crown glass rods (Kiyohara Optics), with a diameter of 2 mm.
The homogeneity of the magnetic field along the coil axis within $\pm$1 mm of the center was measured to be approximately within 0.5 \% at a moment of the peak field.\cite{Nakamura2013}
The length of the sample rod was determined to an accuracy within 1 $\mu$m, and the pickup coil was wound around the rod.
We used a polyamide-imide enameled copper wire (AIW wire, TOTOKU Electric Co.), of diameter 0.06 mm for the pickup coil as it is known to have a high resistance to insulation breakdown.
A quartz rod was firmly held by kapton and bakelite tubes in the center of the coil.
A super-insulation foil was wound outside the bakelite tube to protect against significant electromagnetic noise and the flushing light from the imploding liner.
As a result, the outer diameter of the entire sample holder became about 3.5 mm.

Semiconductor lasers (coherent "CUBE") with a wavelength of 404 nm or 638 nm were used as the light source.
Linearly polarized light was transmitted through a sample rod and divided into s- and p-polarized light ($V_s, V_p$) by a Wollaston prism.
The data acquisition was performed in an electromagnetically shielded room, which was several meters away from the anti-explosion chamber.
The polarized light was transferred by optical fibers, and transformed to an electric signal by an O/E transformer (New Focus, 125-MHz Photoreceivers model 1801), then measured by a transient digital recorder (SONY Tektronix, RTD 710A digitizer), or an A/D converter board (Spectrum, M3i.4142-exp). 

%Fujikura GC.200/250

\begin{table*}
\caption{\label{tab:table}Experimental conditions for each setup (\#F1 to \#C2). The material, sample length $L$, laser wavelength used for FR measurements $\lambda$, total energy, charged voltage, and capacitance of the main condenser bank are given.}
\begin{center}
\begin{tabular}{c|ccc|ccc}
 \hline
 \hline
& \multicolumn{3}{c|}{FR condition} & \multicolumn{3}{c}{Main bank condition} \\
 Exp. & material & $L$ [mm] & $\lambda$ [nm] & Main energy [MJ] & voltage [kV] & capacitance [mF]  \\
 \hline
 %2012/05/24, H #Ato\F1
\#F1 & fused quartz & 2.023 & 404 & 3.5 & 35 & 5.625 (slow)  \\
 %2011/11/24, F (Vs, rev. Vs) #Cto#F2
\#F2 & fused quartz & 2.023 & 404 & 4.0 & 40 & 5.0 (fast)  \\
 %2011/11/11, E #Dto#F3
\#F3 & fused quartz & 0.618 & 638 & 4.0 & 40 & 5.0 (fast)  \\
 %2011/12/01, G
%\#E & fused quartz & 2.023 & 404 & 4.5 & 40 & 5.625  \\
 %2011/06/01, I #Bto#C1
 %verdet: 0.81, B*0.97
\#C1 & crown glass & 1.999 & 404 & 3.5 & 35 & 5.625 (slow)  \\
 %2012/08/17, K #Eto#C2
 %verdet: 0.78*1.04=0.811, B*0.95
\#C2 & crown glass & 1.999 & 404 & 4.0 & 40 & 5.0 (fast)  \\
 \hline
 \hline
\end{tabular}
\end{center}
\end{table*}

%(a) location: C:\測定データ\Faraday回転まとめ\Faraday_all.pxp\Graph1_4,Graph1_1
%(b) location: C:\測定データ\Faraday回転まとめ\Faraday_all.pxp\Graph1_5,Graph1_6
\begin{figure}
\includegraphics[width=4.7cm]{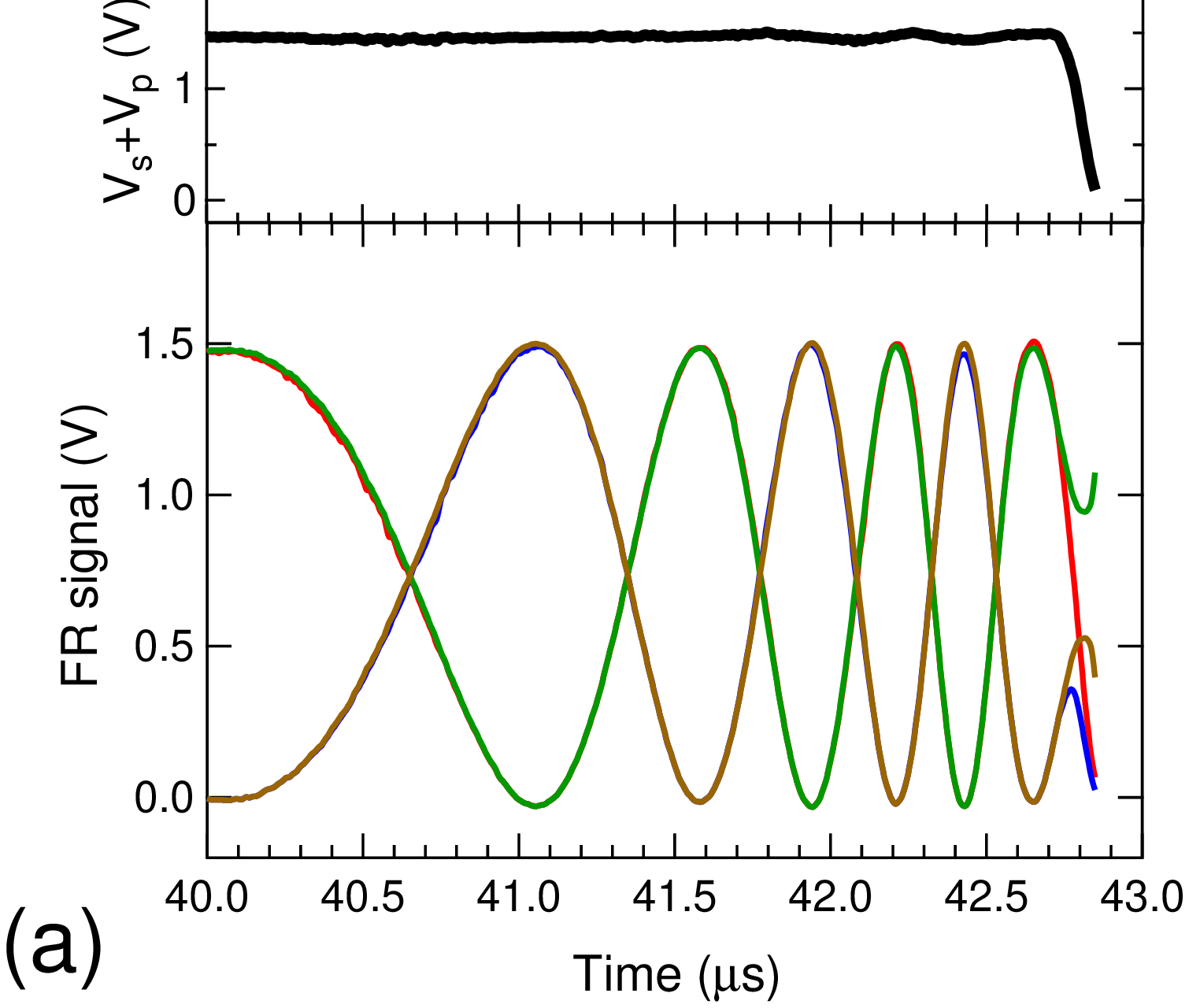}
\includegraphics[width=3.6cm]{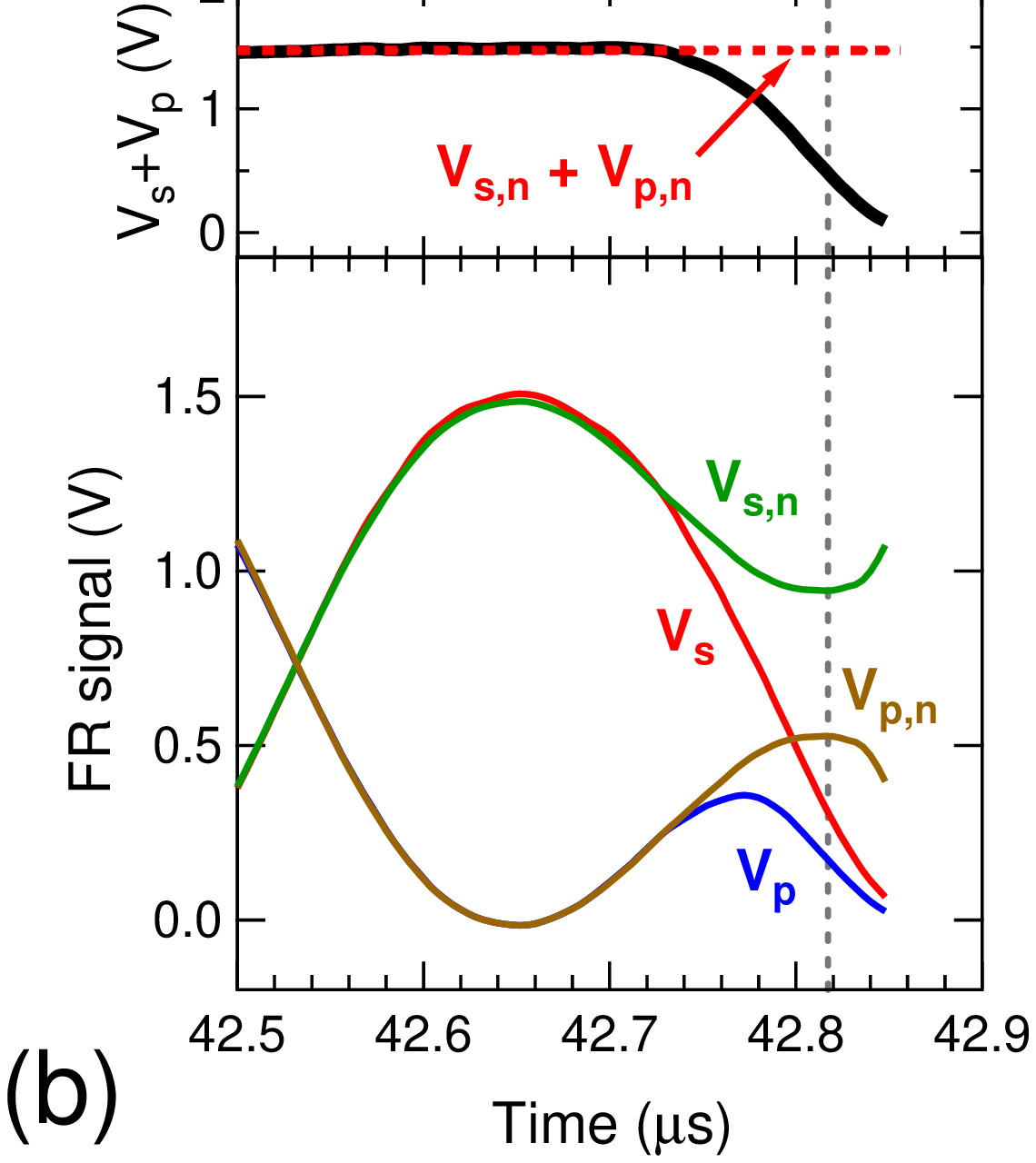}
\caption{\label{fig:3} (color online) (a) The FR signal of the s- and p-polarized components ($V_s$, $V_p$), in experiment \#F1.
The upper panel shows the sum of the two amplitudes of polarized light, $V_s + V_p$.
(b) A plot close to the turn-around phenomenon showing both the raw and normalized data ($V_s$, $V_p$, $V_{s,n}$, and $V_{p,n}$).
The dashed line in the upper panel shows $V_{s,n} + V_{p,n}$.}
\end{figure}

The raw and normalized FR signals ($V_{s}$, $V_{p}$, $V_{s,n}$, and $V_{p,n}$, respectively) are plotted in Fig. 3.
Fig. 3(b) is an enlarged plot about the "turn-around" point.
Except for the final stage of the liner implosion, an almost constant optical transmission signal was obtained (within a 1 \% fluctuation).
However, at the final stage of the liner implosion, when the liner approached the sample holder, an abrupt change of the transmittance occurred, possibly arising from disturbance of the optical pass due to movement of the sample holder.
The total transmittance $V_s + V_p$, shown in the upper panel of Fig. 3 started to decrease from approximately 42.7 $\mu$s, therefore, we normalized the raw FR signal by dividing through by $V_s + V_p$.
The turn-around phenomenon is clearly demonstrated at 42.8 $\mu$s in the FR signal after the normalization procedure, as shown in Fig. 3(b).
The observation of the turn-around feature is important for reliable calibration of the pickup coil to the end point of the flux compression.
Hereafter, we are only concerned with the normalized data for the FR signals.
$\theta_F$ is calculated using $V_s$ and $V_p$ as: 
\begin{equation}
 \theta \textrm{[deg.]} = \frac{180}{2\pi} \times \arccos \left( \frac{V_s - V_p}{V_s + V_p} \right).
\end{equation}

A pickup coil with a 75 cm long twisted copper wire was connected to a 40 m long BNC cable (RG58C/U), in a double-shielded box (as shown in Fig. 2), and monitored by a transient digital recorder in a shielded room separate from the explosion chamber.
As the induced voltage in the pickup coils reached a maximum of a few kV, the signal was transferred into two branches: a hand-made -34 dB attenuator, and a CR integrator (with time constant $\sim$ 1 ms).
After data acquisition, the magnetic field measured by the pickup coil was calculated using the following formula:

%Fig:4 pickup coil
%location: C:\測定データ\Faraday回転まとめ\Faraday_all.pxp\Graph0_4
\begin{figure}
\includegraphics[width=0.8\linewidth]{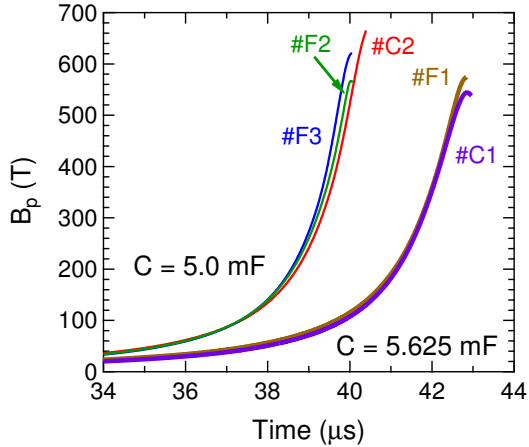}
\caption{\label{fig:4} (color online) The magnetic field curves obtained by the pickup coil, plotted as a function of time. The experimental conditions of \#F1 to \#C2 are summarized in Table I.}
\end{figure}

\begin{equation}
 B = 
\begin{cases}
  &-gS^{-1} \frac{R_L+R_M}{R_M} \int V_3(t) dt \ \ \ \text{: attenuator} \\
  &-(SRC)^{-1} \frac{R_L+R_M}{R_M} V_4(t) \ \ \ \text{: CR integrator} 
\end{cases}
\end{equation}
where, $g$ is a ratio of the attenuator; $R_L$ and $R_M$ are the resistances of the transmission line and the matching circuit (shown in Fig. 2); $RC$ is the time constant of the integrator circuit; $V_3$ and $V_4$ are the measured voltages at the transient recorder (after the attenuator), and the CR integrator, respectively; and $S$ is the cross-sectional area of the pickup coils, calibrated by comparing the induced voltage by an AC magnetic field ($f$ = 50 kHz) with a standard coil.
We discuss the response of the electronic circuit in more detail later.

In Table I we summarize the experimental conditions employed in this article, where the seed field was 3.8 T.
The experimental parameters in Table I were chosen for comparison in an attempt to clarify the imperfection of measurements using the pickup coils.
Different materials for the FR measurement were used to check whether there were material-dependent contributions.
In addition, as described in the introduction, a difference in the high frequency characteristics of the pickup coil signal should be carefully examined.
For this reason, we performed the EMFC experiments with different time profiles by varying the capacitance in the main condenser bank.
Although the implosion speed of the liner is not only a function of capacitance in the main condenser bank, for convenience we separated the dynamics of the liner to either "slow" or "fast" depending on the magnitude of the capacitance.

\section{\label{sec:level3}Results and Discussion}

In Figure 4, we summarize the magnetic field calculated by Eq. (3) from the pickup coil signal ($B_p$), in experiments \#F1 to \#C2.
A "time zero" was determined by the appearance of the discharging trigger noise in the pickup coil signal.
In all experiments, a slowdown of the magnetic field curve was noticed near the peak, which is indicative of the turn-around phenomenon.
The peak of the magnetic field was reached more slowly in \#F1 and \#C1 than in \#F2, \#F3, and \#C2, due to the difference in capacitance of the main condenser bank. 

\subsection{\label{sec:level3-1}Fused quartz measurements}

%Fig.5: #A, #C: FR and pickup coil result
%(a) location: C:\測定データ\Faraday回転まとめ\Faraday_all.pxp\Graph5_4
\begin{figure}
\includegraphics[width=0.75\linewidth]{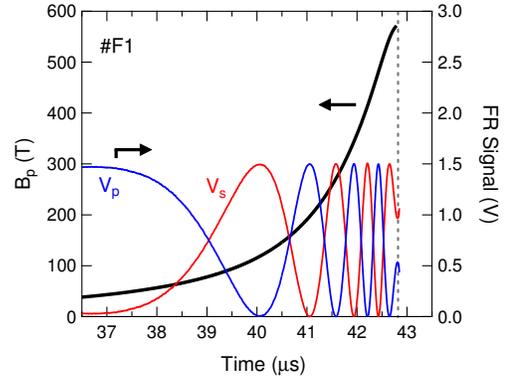}
\caption{\label{fig:5} (color online) The simultaneous measurement of the magnetic field by the pickup coils, and the FR angle of fused quartz, for experiment \#F1.
The thick line corresponds to the magnetic field measured by the pickup coils, and the thin lines correspond to the amplitude of the s- ($V_s$) and p-polarized ($V_p$) FR signals.
The dotted vertical line indicates the time of the turn-around phenomenon observed in the FR signal.}
\end{figure}

First, we discuss the FR results of fused quartz for experiments using different values of the capacitance (of the main condenser bank), \#F1 and \#F2.
The simultaneous measurement of the induced voltage in the pickup coil and the FR angle of the fused quartz, was performed to high precision.
Figure 5 shows a typical result of experiment \#F1, where the thick line corresponds to $B_p$ and the thin lines correspond to $V_s$, and $V_p$.
Note that a turn-around phenomenon was observed in the FR signal (dotted line), the first time that a turn-around structure above 500 T has been clearly observed in an optical signal. 

%Fig.6: #A,#C
%-----------------------------------------
%The results of the comparative experiments
%location: C:\測定データ\Faraday回転まとめ\Faraday_all.pxp\Graph9_11
\begin{figure}
\includegraphics[width=7cm]{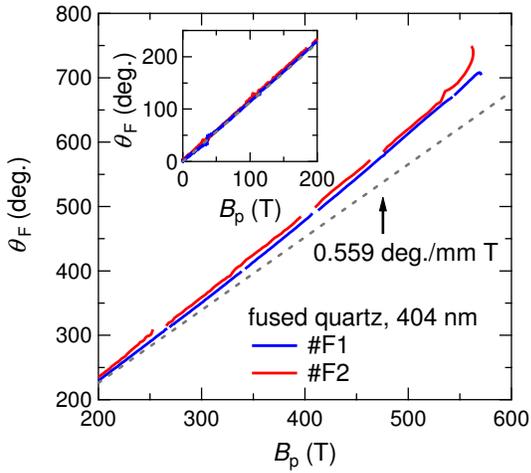}
\caption{\label{fig:6} (color online) The FR angles $\theta_F$ of fused quartz at $\lambda$ = 404 nm, as a function of magnetic field measured by the pickup coils, $B_p$.
The dotted line corresponds to the linear fit of $\theta_F$ for magnetic fields less than 200 T (inset).}
\end{figure}

%Fig. 7: 
%location: C:\測定データ\Faraday回転まとめ\Faraday_all.pxp\Graph17_4
\begin{figure}
\includegraphics[width=7cm]{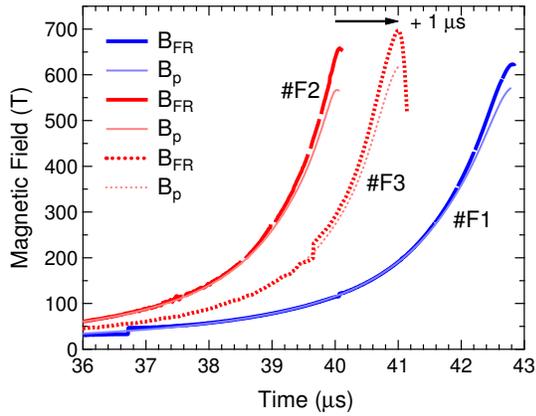}
\caption{\label{fig:7} (color online) The measured magnetic field as a function of time for the induced voltage of the pickup coils (thin lines, $B_p$), and $\theta_F$ of fused quartz (thick lines, $B_{FR}$).
For clarity, the curve corresponding to experiment \#F3 was shifted by 1.0 $\mu$s.}
\end{figure}

%Fig.8: equivalent circuit
%location: C:\測定データ\周波数特性\pickup_Freq_20121120.pptx
\begin{figure*}[htbp]
\includegraphics[width=14cm]{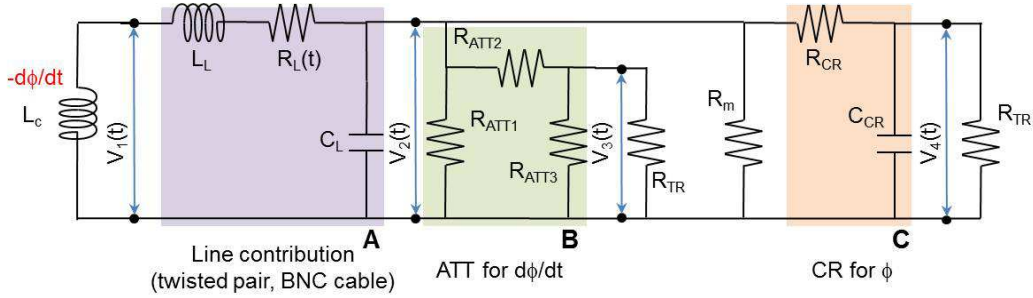}
\caption{\label{fig:8} (color online) The equivalent electric circuit for a measurement of the magnetic field by the pickup coils.
The circuit parameters were as follows: $R_L$ $\sim$ 10 $\Omega$, $R_{ATT1}$ = 100 $\Omega$, $R_{ATT2}$ = 5 k$\Omega$, $R_{ATT3}$ = 100 $\Omega$, $R_m$ = 100 $\Omega$, $R_{CR}$ = 10 k$\Omega$, $C_{CR}$ = 100 nF, and $R_{TR}$ = 1 M$\Omega$.}
\end{figure*}

%Fig.9: constant CLRL
%location: C:\測定データ\Faraday回転まとめ\Faraday_all.pxp\Graph14_6
\begin{figure}[htbp]
\includegraphics[width=6cm]{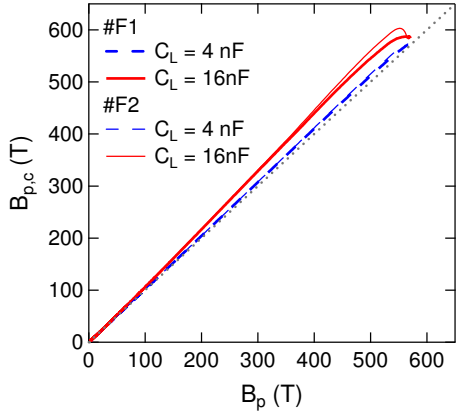}
\caption{\label{fig:9} (color online) The calibrated magnetic field $B_{p,c}$, as a function of $B_p$ for experiments \#F1 and \#F2, respectively, where $C_L$ = 4 nF (dashed lines) and 16 nF (solid lines). The dotted line indicates  $B_{p,c} = B_p$.}
\end{figure}

We derived $\theta_F$ by Eq. (2) and show this as a function of $B_p$ in Fig. 6, where the solid lines correspond to $\theta_F$ and the dotted line corresponds to the linear fit for $B_p <$ 200 T (shown in the inset of Fig. 6).
Below 200 T, the $\theta_F$ of experiments \#F1 and \#F2 coincide well with each other, and respond linearly with respect to $B_p$.
From the slope of the linear fit in Fig. 6, we obtained a Verdet constant of 0.559 $\pm$ 0.008 deg./mm T.
On the other hand, above 200 T a super-linear tendency was observed.
This indicates an increased deviation of $B_p$ from that obtained from the FR angle ($B_{FR}$) on further increase of the magnetic field, which does not relate to the estimate of the Verdet constant for $B_{FR} < $ 200 T.

$B_{FR}$ was calculated from $\theta_F$, assuming that Eq. (1) still holds at high magnetic fields. $B_p$ (thin lines), and $B_{FR}$ (thick lines), are shown alongside one another for comparison in Fig. 7.
For clarity, a horizontal shift of 1.0 $\mu$s was added to the curve corresponding to experiment \#F3.
For experiment \#F3 the Verdet constant was determined to be 0.200 $\pm$ 0.007 deg./mm T for a wavelength of $\lambda$ = 638 nm, which was already confirmed in our previous report.\cite{Nakamura2012JLTP}
A jerky structure in the $B_{FR}(t)$ curve (e.g. see \#F3 at 39.6 $\mu$s), was identified as an artifact.
A discontinuity in the $\theta_F$ curve sometimes occurs as $V_s$ or $V_p$ approach zero, however, the overall FR angle was not affected by this jerky structure in the determination of $B_{FR}$.

$B_{FR}$ was larger than $B_p$ in all experiments, with a maximum ratio [max($B_{FR}$)/max($B_p$)] of 1.08, 1.16, and 1.18 for \#F1, \#F2, and \#F3, respectively.
A similar ratio for experiments \#F2 ($\lambda$ = 404 nm) and \#F3 ($\lambda$ = 638 nm) suggests that the wavelength dependence of the refractive index of the material is not the main source of this difference.
It appears that the "max($B_{FR}$)/max($B_p$)" ratio depends on the capacitance of the condenser bank (\#F1: 5.625 mF; \#F2, \#F3: 5.0 mF), and was lower in \#F1 compared to \#F2 and \#F3, indicating that the rise time of the magnetic field curve contributes to the degree of discrepancy.
These facts suggest that the difference between $B_{FR}$ and $B_p$ majorly arises from the high frequency response of the electric circuit, which has so far not been accounted for.
%#A(9 bank): 618.98/571.49 = 1.08, #C(8 bank): 659.2/566.74 = 1.16, #D(8 bank): 729.47*0.95/620.18 = 1.12

\subsection{\label{sec:level3-B}Analysis: The high frequency response of the circuit}

We analyzed the high-frequency response of the electric circuit used for the pickup coil measurement in our EMFC system.
The equivalent circuit diagram of the signal transmission line from the pickup coil to the transient recorder in Fig. 2 is illustrated in Fig. 8, where the circuit parameters are listed in the caption.
$R_L$ depends on the length of the twisted copper wire used for the pickup coil and is measured for each experiment.
The equivalent circuit is composed of a pickup coil inductance ($L_c$), transmission line (hatched region A), a $\pi$-type attenuator for $d\phi/dt$ measurements (hatched region B), a matching resistance ($R_m$), and a CR integrator for $\phi$ measurements (hatched region C).
$R_{TR}$ is the input impedance of the transient recorder.
The measured matching resistance $R_M$, in Fig. 2, is the combined resistance of the attenuator, the matching resistance $R_m$, and the CR integrator; typically $R_M \sim 50 \ \Omega$.
The measured quantities in the EMFC experiment were $V_3(t)$ and $V_4(t)$, which were used to calculate $V_1(t) = -d\phi/dt$.
In the conventional pulse magnet experiments, $V_1(t)$ is calculated from a simple formula [$V_2(t)(R_L+R_M)/R_M$, see also Eq. (3)], that only takes the resistance ratio into account.
This simplification needs to be revised, however, when $V_2(t)$ contains high frequency components.
The difference between the $B_{FR}$ and $B_p$ discussed in Figs. 6 and 7 should be accounted for well by the high-frequency response of the transmission line with a substantial contribution from $C_L$ and $L_L$.
In the following discussion, the heating effect of the pickup coil during the magnetic field generation is neglected.

The detailed expression of $V_1(t)$ as a function of $V_3 (t)$ [or $V_4 (t)$], is summarized in Appendix A. 
Of this expression, at high frequencies a dominant term $V_1^+(t)$, is described as follows:
\begin{equation}
 V_1^+ (t) =
\begin{cases}& C_L R_L\times R_{\text{CR}}C_{\text{CR}} \frac{d^2V_4 (t)}{dt^2} \\
 & C_L R_L \times \left( 1 + \frac{R_{ATT2}}{R_{ATT3}} + \frac{R_{ATT2}}{R_{TR}} \right) \frac{dV_3 (t)}{dt}
\end{cases}
\label{eq:addterm}
\end{equation}
Since the $L_L$ term appears only in higher order terms, we ignore it in the following discussion. 

At first, we assume that $C_L$ and $R_L$ are constant.
By adding $V_1^+ (t)$ to Eq. (3), we calculated the calibrated magnetic field ($B_{p,c}$), and compare it with $B_p$ in Fig. 9.
The results shown are for the case where $C_L$ = 4 nF (dashed lines) and 16 nF (solid lines), for experiments \#F1 and \#F2.
The dotted line shows $B_{p,c} = B_p$.
These results highlight the fact that the difference between $B_p$ and $B_{p,c}$ is of the same order of magnitude as that of $B_{FR}$ (typically 10 \% for $C_L$ = 16 nF).
However, $B_{FR}$ showed a superlinear dependence with respect to $B_p$ in Fig. 6, whereas $B_{p,c}(B_p)$ for constant $C_L$ exhibits a sublinear dependence in Fig. 9 above 400 T.

%Fig.10: offline measurement of CLRL.
%location: C:\測定データ\周波数特性\20121024_test\Experiment.pxp\Graph0_2
%location: C:\測定データ\周波数特性\20121024_test\Experiment.pxp\Graph1
\begin{figure}
\includegraphics[width=6cm]{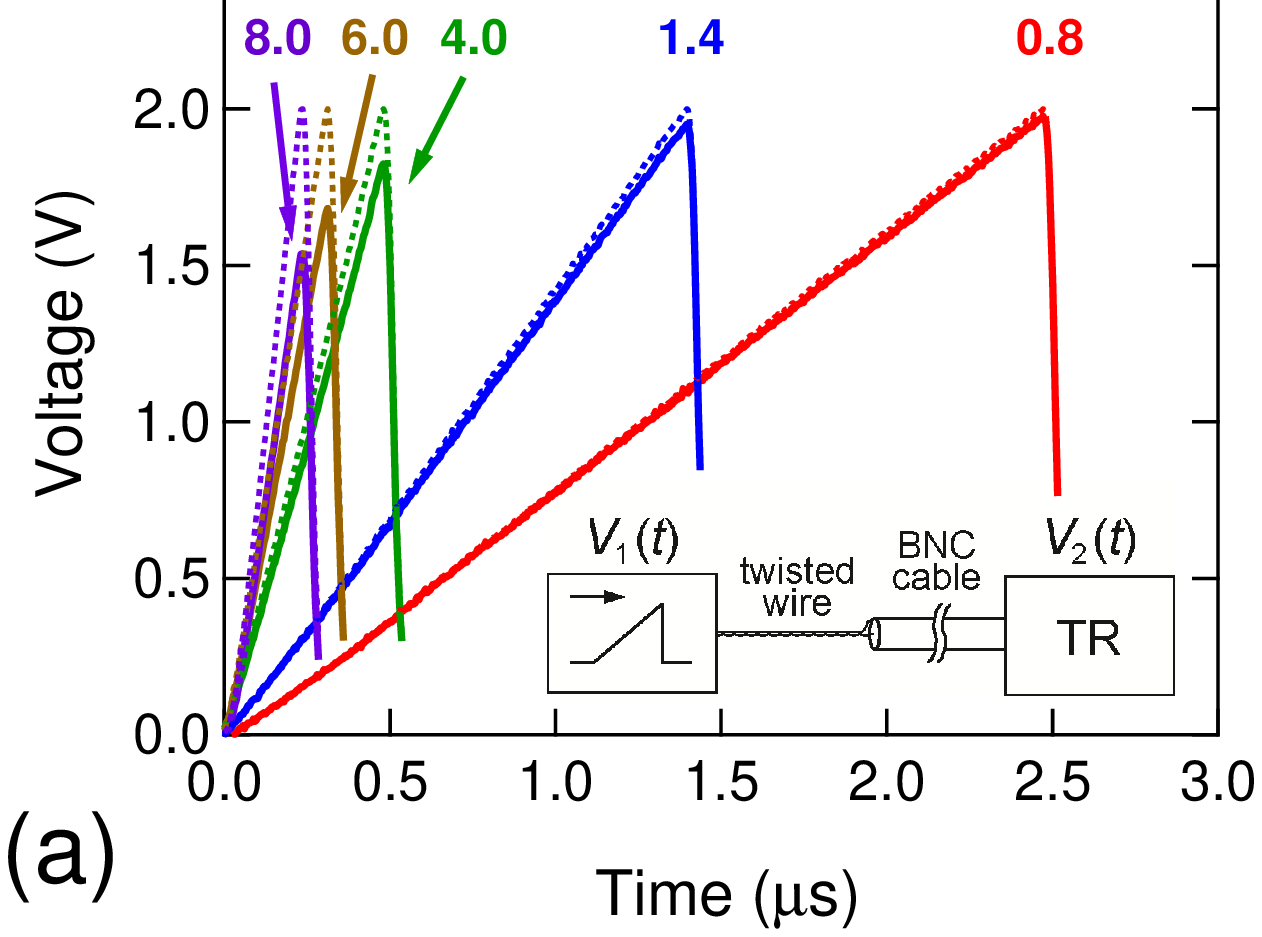}\\
\includegraphics[width=6cm]{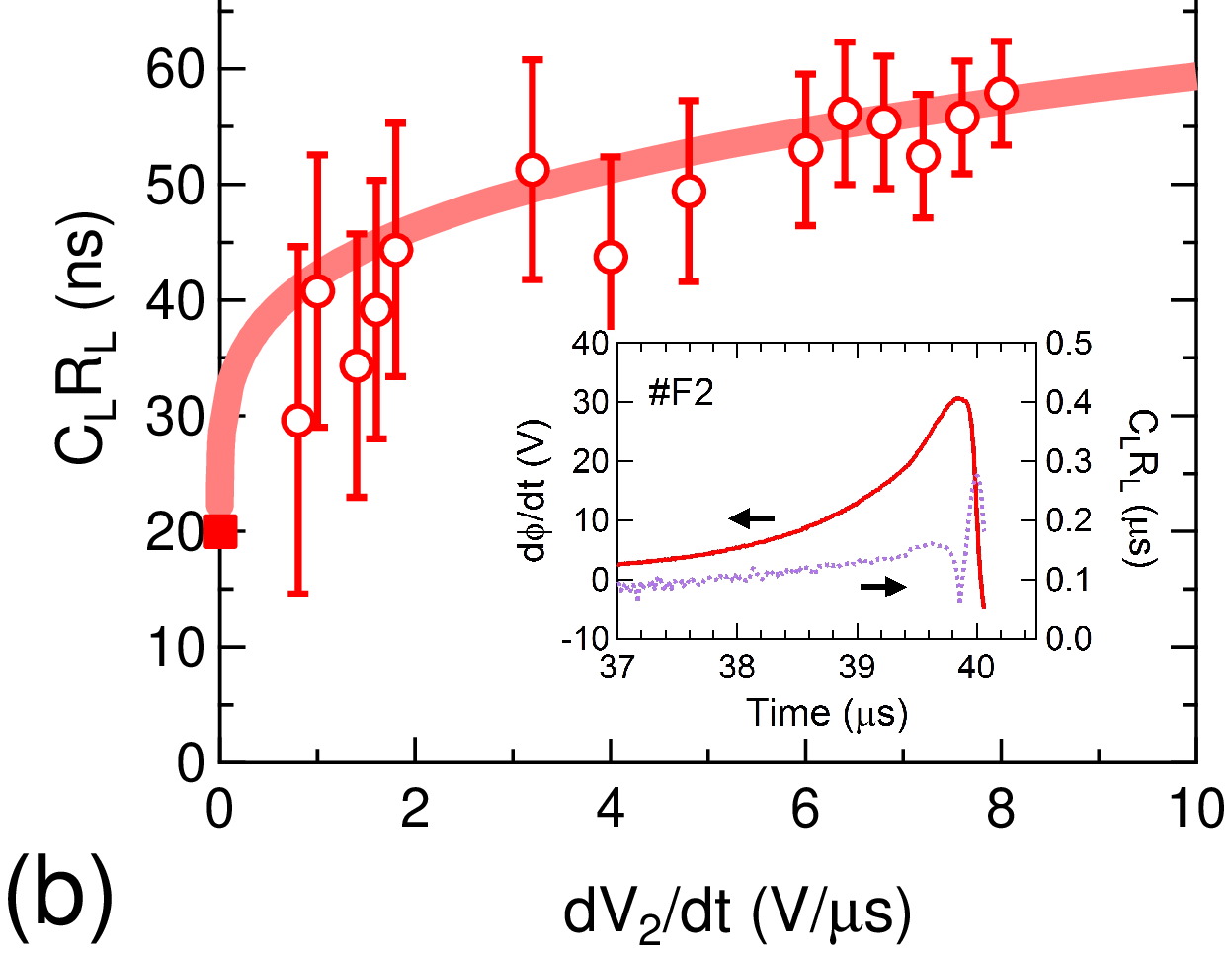}
\caption{\label{fig:10} (color online) (a) The output response of the transmission line obtained by the offline measurement to decide the frequency dependence of $C_L R_L$.
For reference, each input signal is plotted as a dotted line.
The inset shows the experimental setup.
(b) Values of $C_L R_L$ as a function of $dV_2/dt$.
The filled squares are the estimated values under DC conditions; the bold line is the result of fitting.
The inset shows a comparison of $C_L R_L$ and the induced voltage in the pickup coils for experiment \#F2.}
\end{figure}

As a result, we take the frequency dependence of $C_L R_L$ into account.
The signal of the pickup coil $dV_1/dt$, is composed of a wide range of frequencies, therefore to evaluate the empirical formula for $C_L R_L$ of the transmission line in Fig. 8, the transient response of $C_L R_L$ was investigated by using the transmission line separately.
We applied a triangular wave $V_1(t)$, to the open end of a 75 cm long twisted copper wire, as shown in the inset of Fig. 10(a), and measured $V_2(t)$ with respect to $dV_2(t)/dt$ through the same BNC cable used in the EMFC experiment.
In Fig. 10(a), the solid lines correspond to $V_2(t)$ and the dotted lines correspond to $V_1(t)$. Since $V_1(t) \sim V_2 (t) + C_L R_L (dV_2(t)/dt)$ in the hatched region marked A in Fig. 8, we can calculate $C_L R_L$ as a function of $dV_2(t)/dt$ as shown in Fig. 10(b).
The closed square in Fig. 10(b) is the estimated value ($C_L R_L$ = 20 ns) at $dV_2(t)/dt = 0$.
By fitting the data of Fig. 10(b), we obtained an empirical formula for $C_L R_L$ of:
\begin{equation}
C_L R_L = 7 \times 10^{-10} \times |dV_2(t) / dt|^{0.25} + 20 \times 10^{-9}.
\label{eq:CLRL}
\end{equation}
Note that $C_L R_L$ takes a minimum value when $dV_2(t)/dt = 0$.
As an example, $C_L R_L$ and the induced voltage in the pickup coil are compared in the inset of Fig. 10(b).

%Fig. 11a: comparison in case of #H
%Fig. 11b: comparison in case of #F
%Fig. 11a: C:\測定データ\周波数特性\v2_test\20111124_8bank40kV_FR_test2.pxp\Graph4_2
%Fig. 11b: C:\測定データ\周波数特性\v2_test\20120524_9bank35kV_FR_test2.pxp\Graph1_13
%CLRL: 2, line 2  theta:3  B: 1.5  B2: 2, line7
\begin{figure}
\includegraphics[width=8cm]{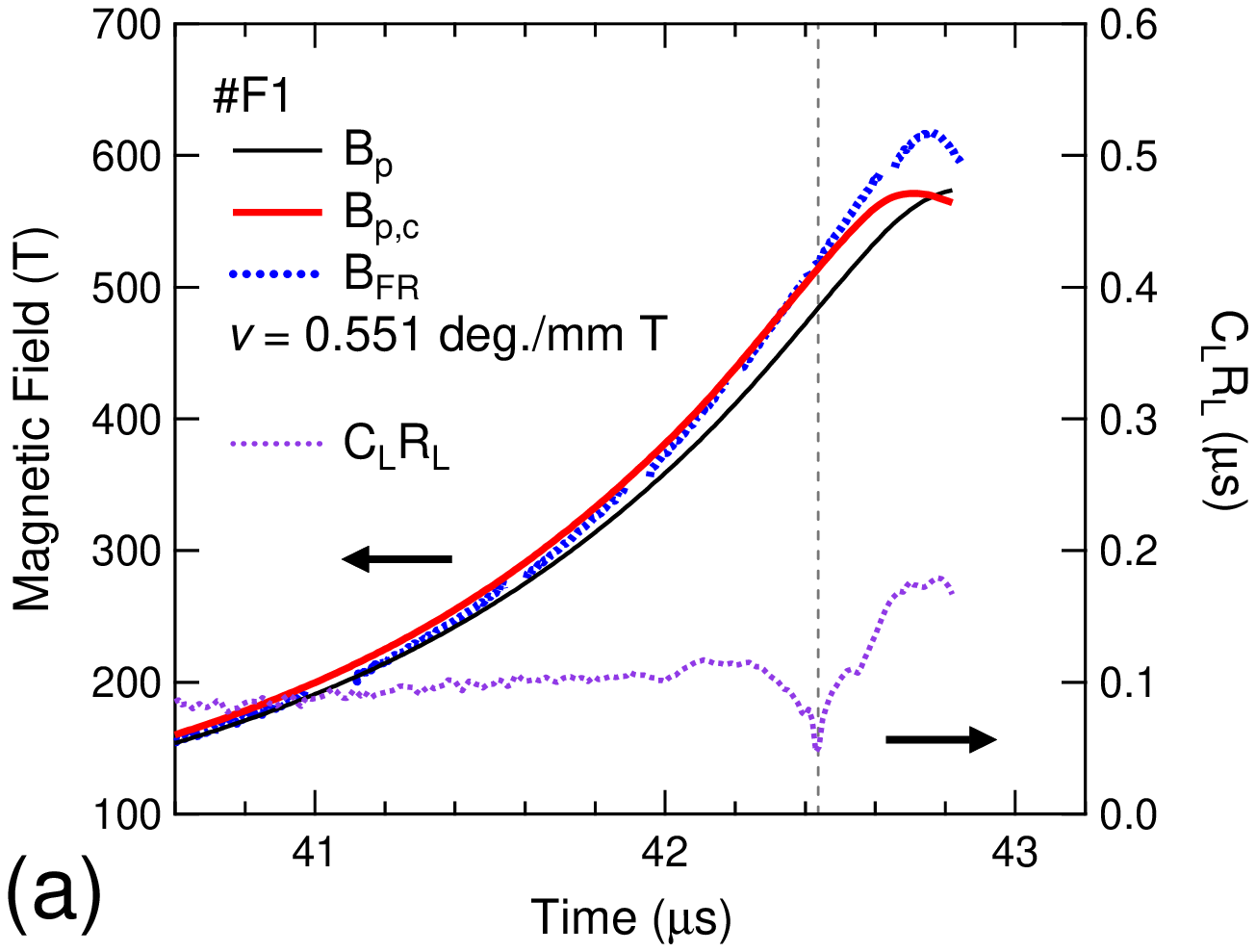}\\
\includegraphics[width=8cm]{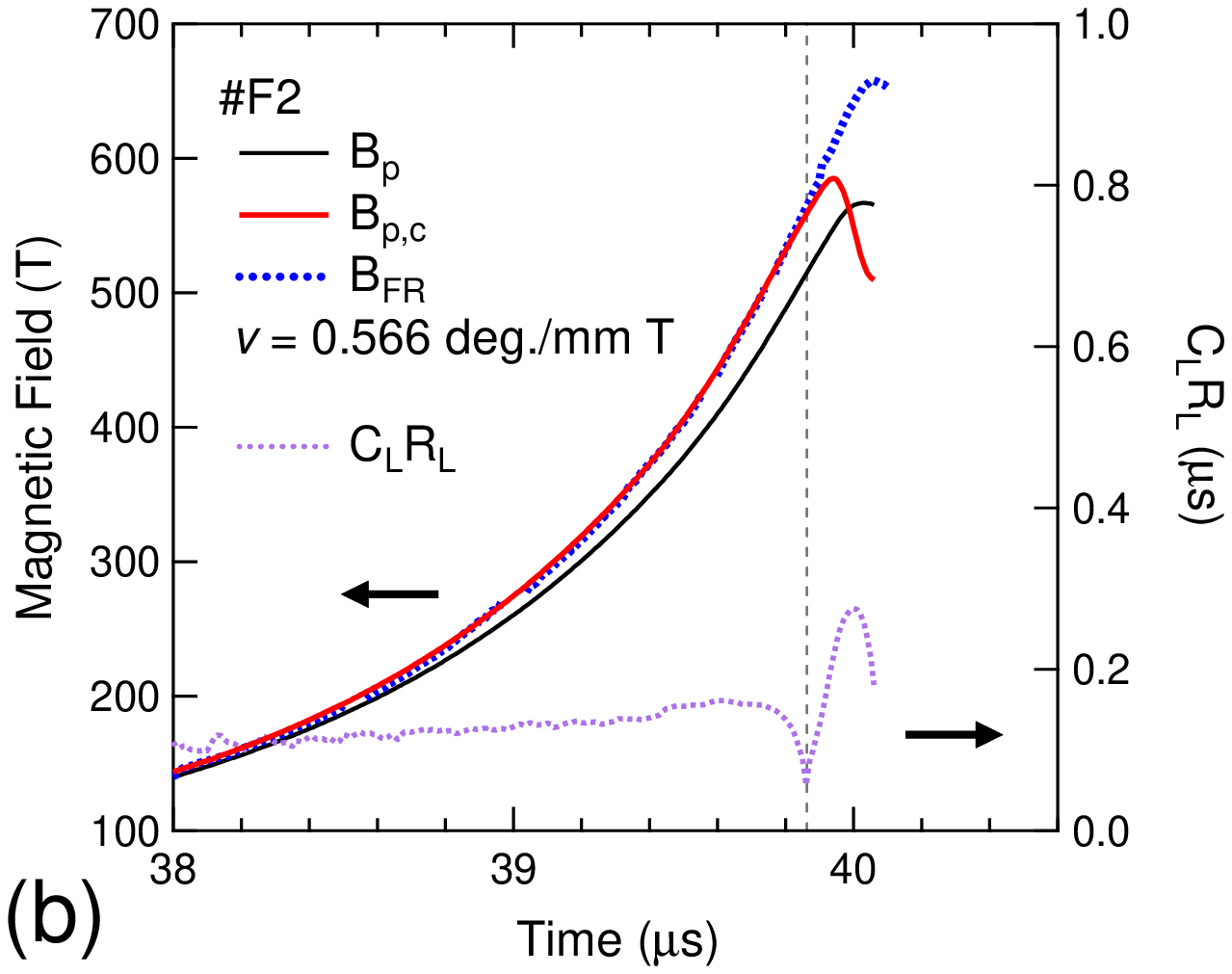}
\caption{\label{fig:11} (color online) The comparison between $B_p$ (solid thin lines), $B_{p,c}$ (solid thick lines), and $B_{FR}$ of fused quartz (thick dashed lines), for experiments (a) \#F1, and (b) \#F2.
The $C_L R_L$ curves are also shown (thin dotted lines), as is the characteristic time of $C_L R_L$ = 0 (vertical dashed line).}
\end{figure}

%-------------------------------------------------
%Fig12: Comparative Experiment
%(a) location: C:\測定データ\周波数特性\v2_test\20120601_9bank35kV_FR_test.pxp\Graph2_7
%(b) location: C:\測定データ\周波数特性\v2_test\20120817_8bank40kV_FR_test.pxp\Graph5_1
\begin{figure*}
\includegraphics[width=8cm]{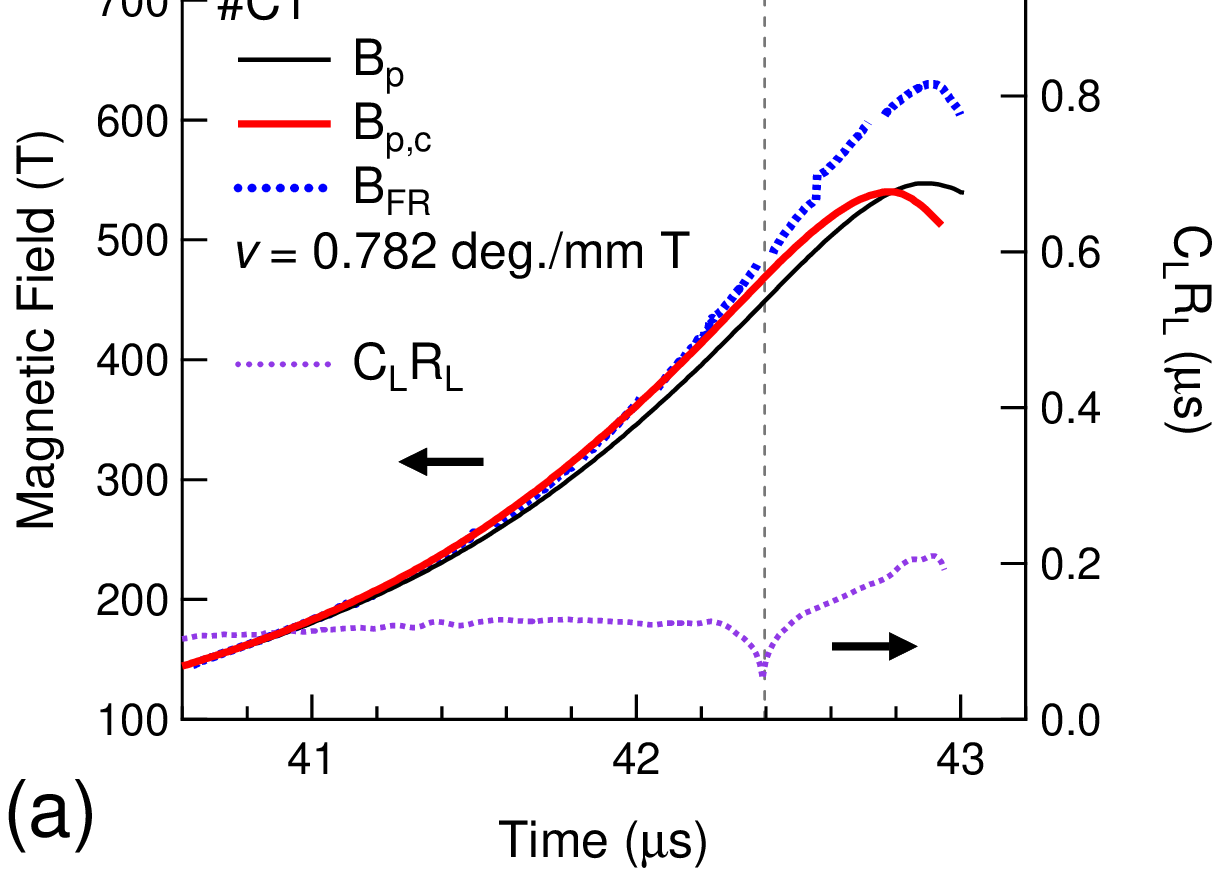}
\includegraphics[width=8cm]{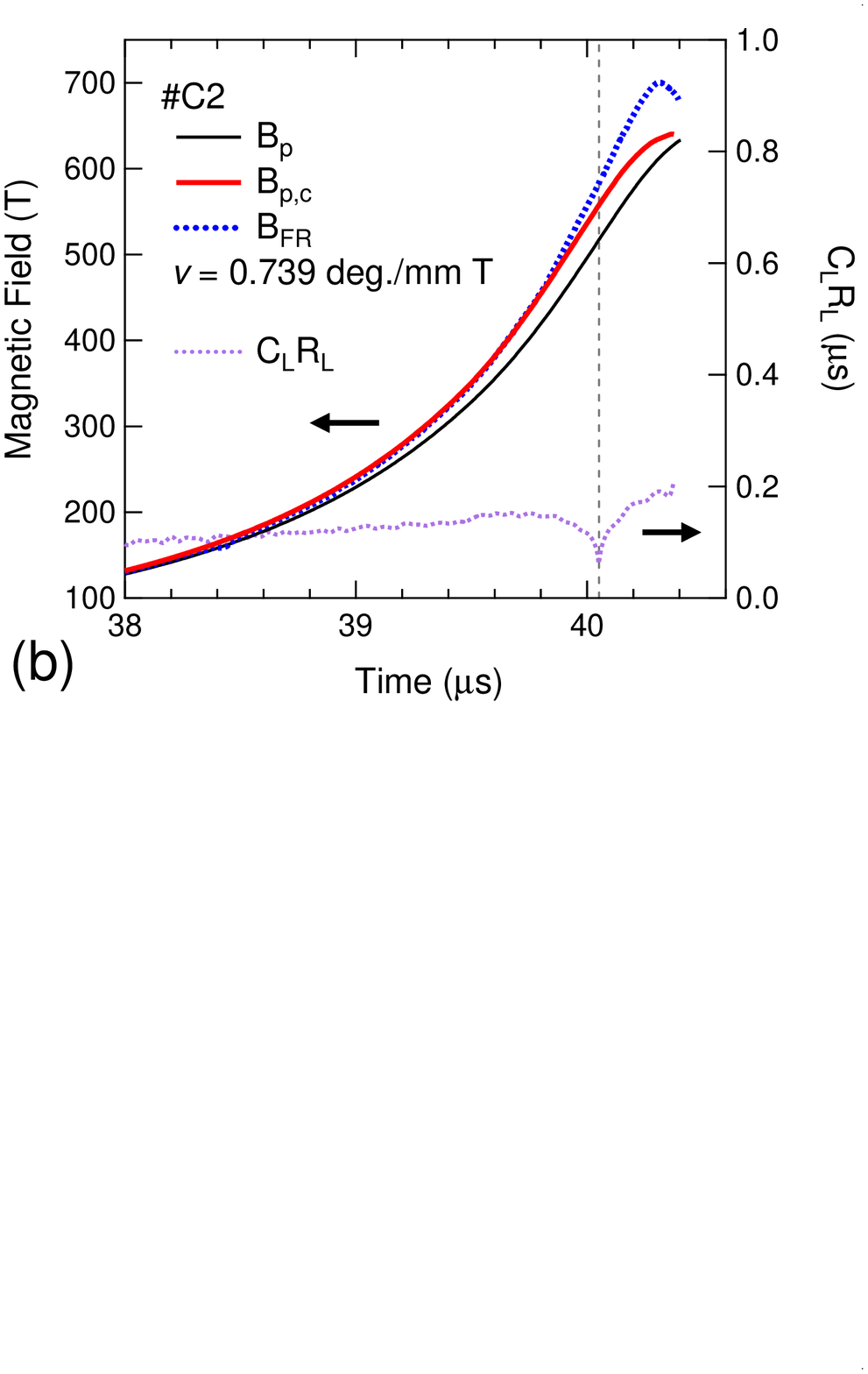}
\caption{\label{fig:12} (color online) The results of simultaneous measurement of the pickup coil and the FR angle of crown glass for experiments (a) \#C1, and (b) \#C2. The notations are the same as for Fig. 11.}
\end{figure*}

By taking into account the frequency-dependent dynamic impedance of the transmission line, $B_p$ is calibrated to coincide with $B_{FR}$.
Using Eqs. (4) and (5), $B_p$ is further calibrated to $B_{p,c}$, and plotted in Fig. 11 for experiments \#F1 (a) and \#F2 (b), where, $B_p$, $B_{p,c}$, $B_{FR}$, $C_L R_L$ correspond to the thin solid line, thick solid line, thick dashed line, and thin dotted line, respectively.
In both cases, $B_{p,c}(t)$ coincides well with $B_{FR}(t)$, exhibiting almost complete overlap up to 500 T for the same parameters, regardless of any difference in their timescales.
Interestingly, $B_{p,c}$ starts to deviate from $B_{FR}$ in both experiments above a characteristic time corresponding to $C_L R_L$ = 0 (vertical dotted lines in Fig. 11).
We discuss this issue further in Sec. III-D.

\subsection{\label{sec:level3-3}Comparative FR measurements using crown glass}

Next, we discuss the results of crown glass as the material for FR measurements.
Crown glass is the generic name for optical glasses with a relatively low refractive index and wavelength dispersion.
Among the many types of crown glass, in this study we used BK7 because of its suitability as a transparent substrate.
Fig. 12 shows the results of simultaneous measurement of the pickup coil and $\theta_F$ of crown glass [(a) \#C1, and (b) \#C2].
The thin and thick solid lines correspond to $B_p$ and $B_{FR}$, respectively.
The Verdet constants used to calculate $B_{FR}$ at 404 nm were 0.782 $\pm$ 0.008 deg./mm T and 0.739 $\pm$ 0.007 deg./mm T in (a) and (b), respectively.
A clear turn-around phenomenon appeared for $B_{FR}$ in both experiments, and in Fig. 12(a) we observed a turn-around phenomenon for both $B_{FR}$ and $B_p$.
The $B_p$ curve close to the turn-around feature, however, seems to be slightly broadened compared to that of $B_{FR}$, suggesting that $B_p$ does not properly reflect the high frequency components.

The calibrated magnetic field of the pickup coil (thick solid line, $B_{p,c}$), was calculated in the same way as in Sec. III-B.
The thin dotted lines of Fig. 12 are the $C_LR_L$ obtained from the analysis using the formula and parameters of Sec. III-B.
A good coincidence between $B_{FR}$ and $B_{p,c}$ was obtained up to the time where $C_LR_L$ = 0 (vertical dotted lines), similar to the case of fused quartz.
These facts indicate that the observed discrepancy between $B_{FR}$ and $B_p$ predominantly arise from the high frequency contributions and not some difference in the materials used for FR.
Furthermore, an additional feature observed for both materials is that $B_p$ cannot be reproduced to the value of $B_{FR}$ beyond the time where $C_LR_L$ = 0.

\subsection{\label{sec:level3-D}Precision of the magnetic field measurement}
%1: discussion about the consistency of Verdet constant 
%      
%2: the discrepancy of magnetic fields (a) - approximately 10 % error, Hugonois plot
%3: the discrepancy of magnetic fields (b) - the limit of a pickup coil
%4: 

We will now consider the precision of measurement of the ultra-high magnetic field generated by the EMFC technique.
The Verdet constant obtained in this study is validated in Appendix B, where it is shown in Table II that the deviation of the Verdet constant ($\delta v/v$) was approximately 1.5 \% for fused quartz and 3 \% for crown glass at 404 nm.
In other words, the precision of the magnetic field evaluated by $\theta_F$ is 3 \% at worst.
On the other hand, we can estimate $\delta v$ using the error in $\theta_F$, $L$, and magnetic field homogeneity.
Since the fluctuation of the optical transmission signal is typically 1 \% (as described in Sec. II), the error in $\theta_F$ ($\delta \theta$), can be estimated using Eq. (2) when $V_s = V_p$, as:
\begin{equation}
 \delta \theta = \frac{180}{\pi}\frac{\delta V_s + \delta V_p}{V_s + V_p} = 0.57 \text{ deg.}
\end{equation}

For the linear fit of Fig. 6, we did not use the data at $V_s$ = 0 or $V_p$ = 0 due to a divergence of the error, therefore, we can assume $\delta \theta \sim$ 0.57 deg.
In Fig. 6, we measured $\theta_F$ up to $\sim$ 700 deg., which would result in $\delta \theta/\theta = 8.14\times 10^{-4}$.
The error due to uncertainty of the exact length of the sample rod $\delta L$, is only 1 $\mu$m, which results in $\delta L/L = 0.5\times10^{-3}$.
For the magnetic field homogeneity along the coil axis, we can assume that $\delta B/B \sim 0.5\times10^{-2}$ within $\pm$ 1 mm of the center of the liner (nearly at the peak field), as described in Sec. II.\cite{Nakamura2013}
As a result, $\delta v/v$ is estimated to be: 
\begin{equation}
 \frac{\delta v}{v} = \sqrt{\left( \frac{\delta \theta}{\theta} \right)^2 +\left( \frac{\delta L}{L} \right)^2 +\left( \frac{\delta B}{B} \right)^2 } = 0.51 \times 10^{-2}.
\end{equation}
The propagation of errors for each experiment therefore results in a total error of the Verdet constant of approximately 1 \%, which is less than the maximum observed deviation of 3 \%.
We therefore suggest that the deviation of the Verdet constant mainly originates from an error in the evaluation of the cross sectional area (i.e. diameter) of the pickup coils.

This result signifies one disadvantage of the pickup coil as a magnetic field probe.
Even if it were possible to correctly calibrate the induced voltage in the pickup coil by the analysis of Sec. III-B, there is still some ambiguity about the precision of the cross-sectional area of the pickup coil itself.
In particular, a diagonal component of the field with respect to the axial direction often induces additional magnetic field in the pickup coil.
Therefore, a precision of less than 3 \% of $B_p$ for the flux compression technique is hardly achievable unless the fabrication of pickup coils is drastically improved.

%Fig. 13: the discussion using the differential value
%(a) location: C:\測定データ\周波数特性\v2_test\diffB.eps\Graph0_5
%(b) location: C:\測定データ\周波数特性\v2_test\diffB.eps\Graph0_3
\begin{figure*}
\includegraphics[width=7cm]{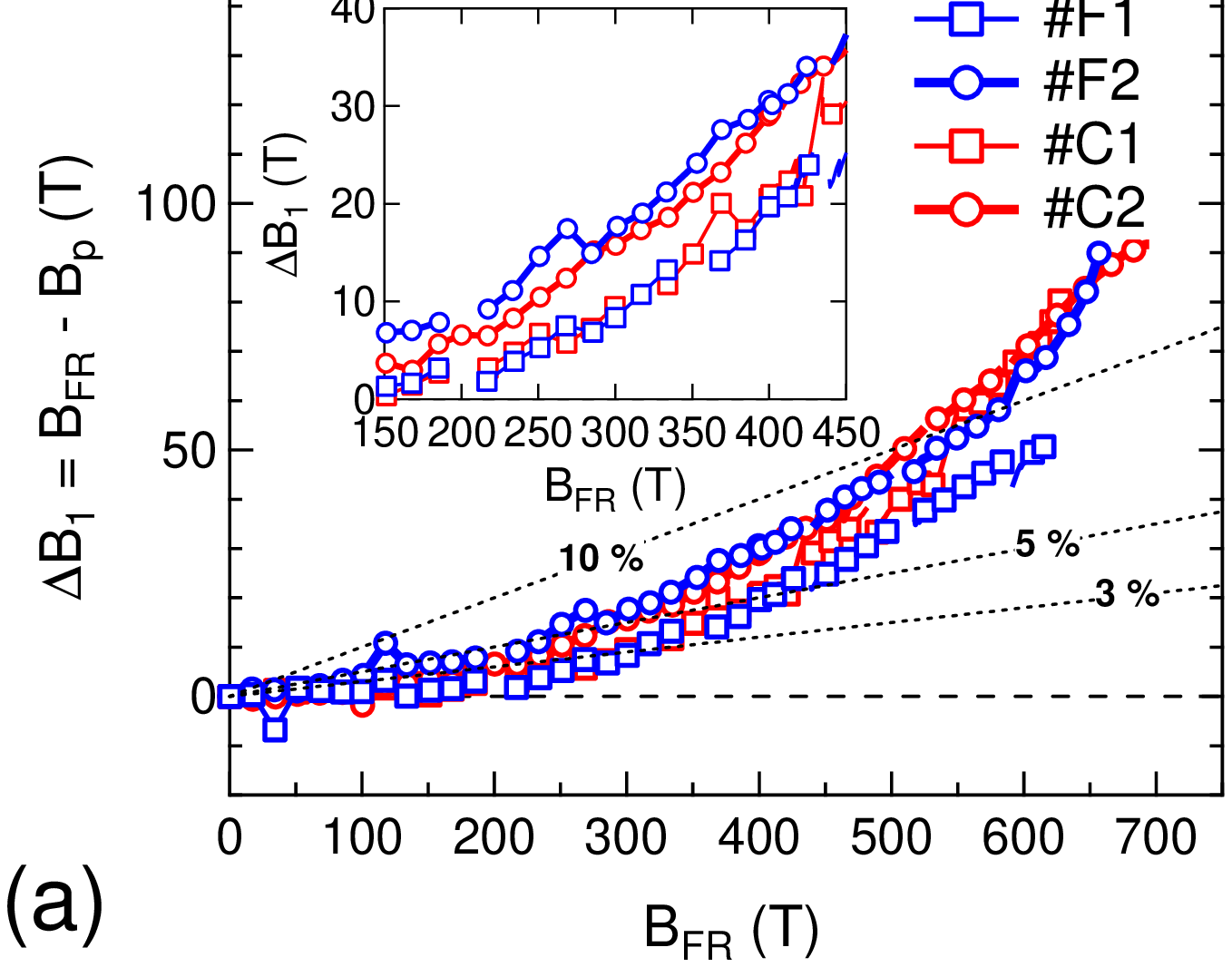}
\includegraphics[width=7cm]{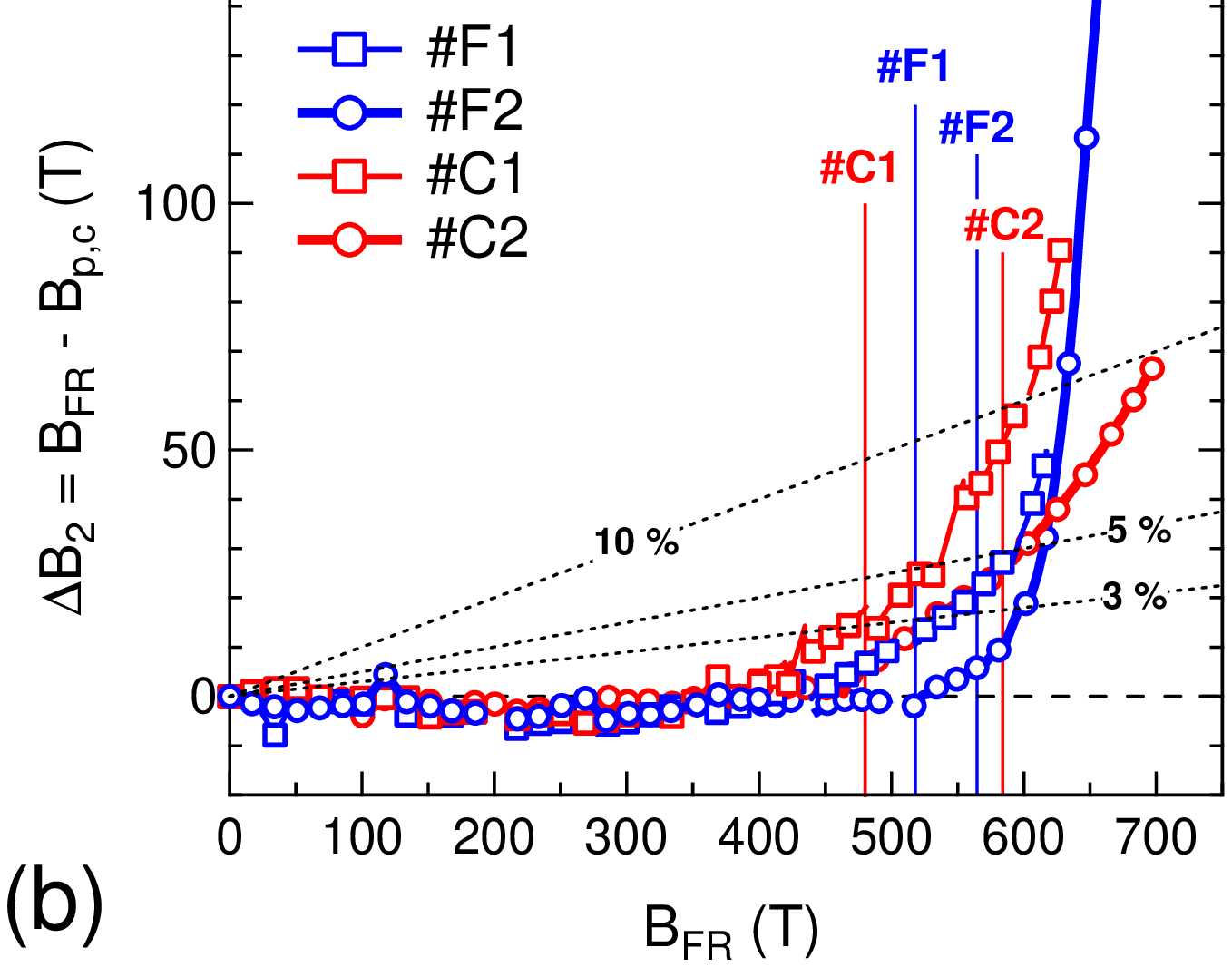}
\caption{\label{fig:13} (color online) Summary of the deviation from $B_{FR}$ for (a) $B_p$, and (b) $B_{p,c}$, plotted against $B_{FR}$.
The dotted lines indicate an estimated error boundary of 3 \%, 5 \%, and 10 \%.
The vertical lines in (b) indicate a position of $B_{FR}$, where the induced voltage in the pickup coil reached a maximum value.
The inset of (a) shows an enlarged plot between 150 T and 450 T.}
\end{figure*}

We will now summarize the discussion about the discrepancy of magnetic fields found in this study.
The mystery of the difference between $B_p$ and $B_{FR}$ was almost solved in the previous section.
The $B_p$ obtained by previous analysis using only a resistance ratio is valid up to magnetic fields of 200 T.
Above 200 T, $B_p$ should be calibrated using the dynamic impedance of the signal transmission line.
Since $B_{FR}$ is always larger than $B_p$, the values of the magnetic field in previous EMFC experiments were likely underestimated.
Fig. 13(a) shows the difference $\Delta B_1 \equiv B_{FR} - B_p$, plotted against $B_{FR}$.
The dotted lines indicate the estimated errors of 3 \%, 5 \%, and 10 \%.
As is evident in the inset of Fig. 13(a), the difference $\Delta B_1$, of experiments with the same capacitance shows almost the same value for different materials.
For example, $\Delta B_1$(400 T) $\sim$ 20 T for \#F1 and \#C1, whilst $\Delta B_1$(400 T) $\sim$ 30 T for \#F2 and \#C2.
Above 200 T $\Delta B_1$ increases suddenly, and for $B_{FR} >$ 600 T a typical $\Delta B_1$ of 10 \% is observed.

In Fig. 13(b) the difference $\Delta B_2 \equiv B_{FR} - B_{p,c}$, is shown as a function of $B_{FR}$.
In all experiments $\Delta B_2$ is almost zero up to 400 T, and increases gradually thereafter (in a different way for each experiment).
The vertical lines indicate a characteristic value of $B_{FR}$ for each experiment, where the induced voltage of the pickup coil takes a maximum value (the point at which $C_LR_L$ approaches zero).
Note that the \#F2 curve behaves quite differently to the others above 600 T, and increases rapidly up to $\Delta B_2$ = 150 T.
This rapid increase of $\Delta B_2$ above 600 T is considered to be due to damage to the pickup coil at high fields, where the induced voltage was extremely high.

Some other possibilities for the origin of the discrepancy in $\Delta B_2$ at very high magnetic fields can also be considered.
The first is an incompleteness of our high frequency analysis.
The equation used to calibrate the induced voltage in the pickup coil was complicated, as shown in Appendix A.
For simplicity we ignored the higher order terms of $C_L$ and $L_L$.
The temporal change in $R_L$ due to heating of the pickup coil during the flux compression step is also an unknown factor (and is considered to occur by eddy currents in the metal), and quantitative evaluation is difficult at present.

Secondly, a pickup coil exposed to a huge induced voltage is damaged to some extent, and therefore becomes incapable of correct measurement of the magnetic field near the turn-around point.
Indeed, $\Delta B_2$ started to increase slightly before $C_LR_L = 0$ [the vertical line in Fig. 13(b)], where the induced voltage was at its maximum value, as shown in the inset of Fig. 10(b).
At the moment where $\Delta B_2$ increases, the induced voltage of the pickup coil becomes as high as an order of 1 kV.

The third possibility is that the assumption used for Eq. (1) does not hold in extremely high magnetic fields.
In this case, $\theta_F$ can be expressed by adding an additional term: $\theta_F = L(vB + f(B))$.
The deviation of $B_p$ from $B_{FR}$ is given by $L f(B)$, which should be time-independent.
However, in Fig. 13(b), $\Delta B_2$ shows different behavior, depending on the choice of the main condenser bank, and exhibited time-dependent behavior rather than material dependence (see \#F1, \#F2 and \#C1, \#C2).
This fact implies that $f(B) \sim 0$ and that the deviation in the linearity of $\theta_F$ in optical glasses does not present a dominant contribution to the discrepancy in $\Delta B_2$ for the highest magnetic fields.
Furthermore, there is difficulty in determining any reason for the deviation from linearity at only the "turn-around" point.

To summarize, the pickup coil method is limited with respect to the precise measurement of the magnetic field in extreme experimental conditions such as the EMFC technique, which is capable of generating magnetic fields above 500 T.
Therefore, the evaluation of the magnetic field using FR measurements together with a pickup coil, is necessary.

\begin{figure*}[!ht]
\begin{minipage}[c]{\textwidth}
\begin{align}
V_1 (t) &= \left\{ R_{\text{CR}}C_{\text{CR}} +R_L \left[ R^{-1}_{\text{ATT-TR}} R_{\text{CR}}C_{\text{CR}} + C_{\text{CR}} (R_{\text{m}}^{-1} R_{\text{CR}} +1) \right] \right\} \frac{d V_4 (t)}{dt} \nonumber \hspace{65mm}\text{(A1)} \\
&+ R_L \Biggl[ \left( R_L^{-1}+R^{-1}_{\text{ATT-TR}} \right) \left( 1+\frac{R_{\text{CR}}}{R_{\text{TR}}} \right) + \left( \frac{1}{R_{\text{m}}} + \frac{R_{\text{m}}^{-1} R_{\text{CR}} +1}{R_{\text{TR}}} \right) \Biggr] V_4 (t)  \nonumber \\
&+ L_L \left\{ \Biggl[ R^{-1}_{\text{ATT-TR}} \left( 1+\frac{R_{\text{CR}}}{R_{\text{TR}}} \right) + \left( \frac{1}{R_{\text{m}}} + \frac{R_{\text{m}}^{-1} R_{\text{CR}} +1}{R_{\text{TR}}} \right) \Biggr] \frac{dV_4 (t)}{dt}  + \left[ R^{-1}_{\text{ATT-TR}} R_{\text{CR}}C_{\text{CR}} + C_{\text{CR}} (R_{\text{m}}^{-1} R_{\text{CR}} +1) \right] \frac{d^2 V_4 (t)}{dt^2} \right\}  \nonumber  \\
&+ C_L R_L \Biggl[ \left( 1+\frac{R_{\text{CR}}}{R_{\text{TR}}} \right) \frac{dV_4 (t)}{dt} + R_{\text{CR}}C_{\text{CR}} \frac{d^2V_4 (t)}{dt^2} \Biggr]  \nonumber \\
&+ C_L L_L \Biggl[ \left( 1+\frac{R_{\text{CR}}}{R_{\text{TR}}} \right) \frac{d^2V_4 (t)}{dt^2} + R_{\text{CR}}C_{\text{CR}} \frac{d^3V_4 (t)}{dt^3} \Biggr],  \nonumber 
\end{align}
\end{minipage}
\end{figure*}

\section{Conclusion}
We succeeded in the precise evaluation of ultra-high magnetic fields of up to 700 T using the Faraday rotation (FR) angle of optical glass, confirmed by the observation of a turn-around structure in the FR signal.
We compared the magnetic fields measured by a pickup coil with that calculated from the FR angle, and found that a deviation starts to appear above 200 T.
As a result of this analysis, for the correct evaluation of the magnetic field measured by a pickup coil, the high frequency response of the signal transmission line must be accounted for in the calibration.
However, this is not sufficient above 500 T.
The precise measurement of ultra-high magnetic fields is only possible by the use of FR measurements of fused quartz or crown glass, in which the linearity of the FR angle was maintained in magnetic fields of up to 700 T.
The values of the magnetic field evaluated by only a pickup coil in previous EMFC experiments were likely underestimated.

\begin{acknowledgements}
We thank Dr. A. Miyata for his support with the experimental setup.
\end{acknowledgements}

\appendix

\section{FORMULIZATION OF THE EQUIVALENT CIRCUIT}

For the circuit in Fig. 8, the induced voltage in the pickup coil $V_1(t)$, can be described as a function of $V_4(t)$ by Eq. (A1).
In Eq. (A1), $R_{\text{ATT-TR}}$ is the combined resistance of the attenuator and the transient recorder: 
\begin{equation}
 R^{-1}_{\text{ATT-TR}} = \frac{R_{\text{ATT1}} + \left( R_{\text{ATT2}} + \frac{R_{\text{ATT3}}R_{\text{TR}}}{R_{\text{ATT3}}+R_{\text{TR}}} \right)}{R_{\text{ATT1}} \times \left( R_{\text{ATT2}} + \frac{R_{\text{ATT3}}R_{\text{TR}}}{R_{\text{ATT3}}+R_{\text{TR}}} \right)}.\nonumber \hspace{5mm}\text{(A2)} 
\end{equation}

\begin{figure*}[tbhp]
\begin{minipage}[c]{\textwidth}
\begin{align}
V_1 (t) &\sim \left\{ \left( 1+\frac{R_{\text{ATT2}}}{R_{\text{ATT3}}}+\frac{R_{\text{ATT2}}}{R_{\text{TR}}} \right) \left[ 1+R_L\left( R_\text{ATT1}^{-1}+R_\text{CR}^{-1}+R_\text{m}^{-1}\right)\right]  \right\}V_3(t) \nonumber \hspace{35mm}\text{(A3)} \\
& + R_L (R_\text{ATT3}^{-1}+R_\text{TR}^{-1}) V_3(t) \nonumber \\
& + L_L \left[ \left( 1+\frac{R_{\text{ATT2}}}{R_{\text{ATT3}}}+\frac{R_{\text{ATT2}}}{R_{\text{TR}}}\right)\left( R_\text{ATT1}^{-1}+R_\text{CR}^{-1}+R_\text{m}^{-1}\right) + R_L(R_\text{ATT3}^{-1}+R_\text{TR}^{-1}) \right] \frac{dV_3 (t)}{dt} \nonumber \\
& + C_L R_L \left(1+\frac{R_{\text{ATT2}}}{R_{\text{ATT3}}}+\frac{R_{\text{ATT2}}}{R_{\text{TR}}}\right) \frac{dV_3 (t)}{dt} \nonumber \\
& + C_L L_L \left(1+\frac{R_{\text{ATT2}}}{R_{\text{ATT3}}}+\frac{R_{\text{ATT2}}}{R_{\text{TR}}}\right) \frac{d^2 V_3 (t)}{dt^2} \nonumber 
\end{align}
\end{minipage}
\end{figure*}

Although it is more complex to express $V_1(t)$ as a function of $V_3(t)$, we can describe it in a similar way by Eq. (A3).
In Eqs. (A1) and (A3), the first term is the first approximation of the circuit response, which has been used before.
The remaining terms (which contain $L_L$ and/or $C_L$), become important at high frequencies.
The dominant term $V_1^+(t)$, as referenced in section III-B, appears in the fourth term of Eqs. (A1) and (A3).

\section{THE WAVELENGTH DISPERSION OF THE VERDET CONSTANT}

We discuss here the consistency of the Verdet constant used to calculate $B_{FR}$.
For the wavelength dispersion of the Verdet constant in an optical glass, the experimental data is known to be well described by the following empirical formula:\cite{Bach1995text}
\begin{equation}
 v = \frac{\pi}{\lambda} \left( a + \frac{b}{\lambda^2 - \lambda_0^2} \right),
\end{equation}
where $a$, $b$, and $\lambda_0$ are fitting parameters.
Fig. 14 shows the $v(\lambda)$ of fused quartz and crown glass in the visible light region.
The closed symbols are the data taken for fused quartz by Garn {\it et al.} \cite{Garn1968} and Ramaseshan,\cite{Ramaseshan1946} and the dashed curves are the curves fitted by Eq. (B1).
The solid curve is the calculation for crown glass using the fitting parameters given in Table II.
The vertical dotted lines show the wavelengths investigated in this study: 404 nm and 638 nm.
In Table II, we listed $a$, $b$, and $\lambda_0$ for each material, and the calculated and experimental Verdet constant ($v_{cal}$, $v_{exp}$), at 404 nm and 638 nm.
The fitting parameters used for fused quartz ($a$, $b$, $\lambda_0$), are the result of fitting by Garn {\it et al.} \cite{Garn1968} and Ramaseshan\cite{Ramaseshan1946} by Eq. (6).
The open symbols in Fig. 14 show $v_{exp}$ determined in this study, which coincides well (within error) with values reported earlier.

%Fig14: Verdet constant of optical glass
%The Properties of Optical Glass, Eds. Hans Bach and Norbert Neuroth, (1995, Springer)
%location: C:\測定データ\Faraday回転まとめ\Garn1968\Graph0_13
\begin{figure}
\includegraphics[width=7cm]{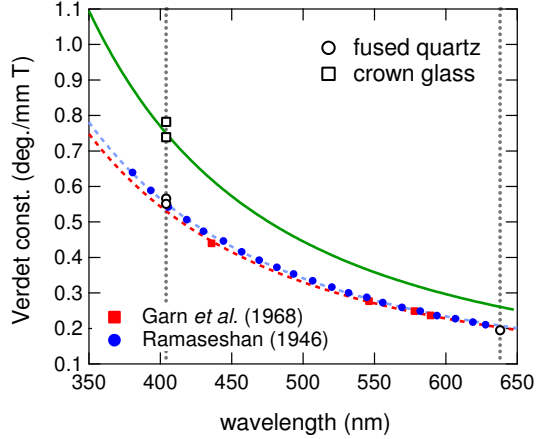}
\caption{\label{fig:14} (color online) The wavelength dispersion of the Verdet constant of fused quartz and crown glass in the visible light region.
The open symbols (circles: fused quartz, squares: crown glass), show the Verdet constants obtained in this study.
The closed symbols are the data taken from Garn {\it et al.} \cite{Garn1968} and Ramaseshan.\cite{Ramaseshan1946}
The dashed curves are the fits to the data of Eq. (B1) in Appendix B.
The theoretical curve of typical crown glass (BK7) was calculated using the fitting parameters in Table II.
The vertical dotted lines show the wavelengths used in this study: 404 nm and 638 nm.}
\end{figure}

\begin{table}
\caption{\label{tab:table}The parameters used for the wavelength dispersion formula [Eq. (B1)] of the Verdet constant in optical glass, and the calculated and experimental Verdet constant ($v_{cal}$, $v_{exp}$) at 404 and 638 nm. }
\begin{center}
\begin{tabular}{c|cc}
 \hline
 \hline
  & fused quartz & crown glass \\
\hline
$a$ [$10^{-9} $ /T] & 400.78\cite{Ramaseshan1946,Bach1995text}, 401.39\cite{Garn1968,Bach1995text} & 489.92\cite{Bach1995text} \\
$b$ [$10^{-20}$ m$^2$/T] & 12.900\cite{Ramaseshan1946,Bach1995text}, 13.671\cite{Garn1968,Bach1995text} & 21.751\cite{Bach1995text} \\
$\lambda_0$ [$10^{-9}$ m] & 0.926\cite{Ramaseshan1946,Bach1995text}, 16.033\cite{Garn1968,Bach1995text} & 101.0\cite{Bach1995text} \\
\hline
$v_{cal}$ (404 nm) & 0.531\cite{Ramaseshan1946,Bach1995text}, 0.552\cite{Garn1968,Bach1995text} & 0.750\cite{Bach1995text} \\
$v_{exp}$ (404 nm) & 0.559 $\pm$ 0.008 & 0.761 $\pm$ 0.022 \\
$v_{cal}$ (638 nm) & 0.203\cite{Ramaseshan1946,Bach1995text}, 0.207\cite{Garn1968,Bach1995text} & 0.261\cite{Bach1995text} \\
$v_{exp}$ (638 nm) & 0.200 $\pm$ 0.007 & -\\
 \hline
 \hline
\end{tabular}
\end{center}
\end{table}

%\nocite{*}
%\bibliography{aipsamp}% Produces the bibliography via BibTeX.

\end{document}